\documentclass[12pt]{article}
\usepackage[pctex32]{graphics}
\begin{document}
\vspace{2cm}
\begin{center}
~
\\
~
{\bf  \Large ADM Mass on General Spacetime and  Hawking-Page Phase Transition in Magnetic Black Brane}
\vspace{1cm}

                      Wung-Hong Huang\\
                       Department of Physics\\
                       National Cheng Kung University\\
                       Tainan,Taiwan\\

\end{center}
\vspace{2cm}
We derive a formula which enable us to evaluate the ADM mass in a more general  curved spacetime. We apply the formula to evaluate the thermodynamical quantities of the Melvin magnetic black D-branes.  We see that there is the Hawking-Page phase transition and the corresponding dual gauge theory will show the confinement-deconfinement transition under magnetic flux.

\vspace{2cm}
\begin{flushleft}
*E-mail:  whhwung@mail.ncku.edu.tw\\
\end{flushleft}
\newpage
\section{Introduction}
 Historically, the formula of ADM mass in the case with metric form
$$ds^2=A(U)\eta_{\mu\nu}dx^\mu dx^\nu+B(U) \delta_{mn}dy^a dy^b, \eqno{(1.1)}$$
in which $U^2\equiv\delta_{mn}dy^a dy^b$  had been first derived [1,2]. Next, Lu [3] had derived a general formula of ADM mass in the case with metric form  
$$ds^2=-A(U)dt^2+B(U)dU^2+C(U)U^2d\Omega_d^2 + D(U)\delta_{ij}dx^i dx^j,  \eqno{(1.2)}$$
which has been wildly used in many literatures since then [4-7]. 

In this paper we want to consider the spacetime associated to the black D-brane under the Melvin magnetic field [8], which is [9]
$$ds_{10}^2 =\sqrt{1+ B^2 r^2} \left[H^{-1\over2}\left(-f(U)~dt^2+dz^2+dw^2+dr^2+ {r^2d\phi^2\over 1+B^2r^2}\right)+H^{1\over2} \left(f(U)~dU^2+U^2d\Omega_4^2\right) \right],\eqno{(1.3)}$$ 
 However, as our metric does not fall in above class we have to derive a slightly general formula to calculate the ADM mass. 

In section II we derive a more general formula which enable us to calculate  the ADM mass in our cases.  In section II we use this formula to evaluate the thermodynamical quantities of the black D-branes with magnetic field, which is dual to the finite temperature gauge theory under the magnetic field. We have found the Hawking-Page transition for sufficiently large  magnetic field.  The last section is devoted to a short conclusion. 
\section{ADM Mass in More General Geometry}
Consider a general black p-brane with metric $g_{MN}= g^{(0)}_{MN}+h_{MN}$ in which $g^{(0)}$ is the D dimensional flat limit of the corresponding space-time metric. $h_{MN}$ is asymptotically zero but not necessarily small everywhere [1-3].  To first order in  $h_{MN}$ the Einstein equation looks like 
$$ R^{(1)}_{MN} -{1\over2} g^{(0)}_{MN} R^{(1)} = \kappa^2 \Theta_{MN}. \eqno{(2.1)}$$
The ADM mass per unit volume is defined as
$$M= \int d^{D-d-1}y ~\Theta_{00}.\eqno{(2.2)}$$
The general $R^{(1)}_{MN}$ has been given in [1,2] as 
$$R^{(1)}_{MN}= {1\over2}\left({\partial^2h^P_{~M}\over \partial x^P\partial x^N}+{\partial^2h^P_{~N}\over \partial x^P\partial x^M}-{\partial^2h^P_{~P}\over \partial x^M\partial x^N}-{\partial^2h_{MN}\over \partial x^P\partial x^P}\right),\eqno{(2.3)}$$
where the indices are raised and lowered using the flat Minkowski metric.  Using (2.1) and (2.3) we find that 
$$  \kappa^2 \Theta_{00}= - {1\over2}{\partial^2h^0_{~0}\over \partial x^Q\partial x_Q}+ {1\over2}{\partial^2h\over \partial x^Q\partial x_Q}+ {1\over2}{\partial^2h^M_{~N}\over \partial x^M\partial x_N},~~~~h\equiv \sum_P h^P_{~P} \eqno{(2.4)}$$

  We will consider the more general metric which has a following block form
$$ds^2= -A(U,r,..)dt^2+\left[B(U,r,..)dU^2+C(U,r,..)U^2d\Omega_{d-1}^2 \right]\hspace{6cm}$$
$$ \left[E(U,r,..)dr^2+F(U,r,..)r^2d\Omega_{D-1}^2 \right] + G(U,r,..)\sum_i ^{d_x}dx_i^2 +\cdot\cdot\cdot, \eqno{(2.5)}$$
In below we present a systematic procedure to find the ADM mass.

$\bullet$ {\bf Step 1 :}  The first property we can see is that the first term in (2.4) will be canceled by the $h^0_{~0}$ term in second term, as $h= h^0_{~0}+\cdot\cdot\cdot$.  Thus we conclude that $ \Theta_{00}$ does not depend on $A$.  

$\bullet$ {\bf Step 2 :}  To see how the $B$ and $C$ will appear in $ \Theta_{00}$ we first rewrite a part of line element in the coordinate as follow
$$BdU^2+CU^2d\Omega_{d-1}^2=(B-C)dU^2+C(dU^2+U^2d\Omega_{d-1}^2)\hspace{3.5cm}$$
$$={B-C\over U^2}\sum_{i=1}^d U_i U_j dU_i dU_j+C\sum_{i=1}^d dU_i^2, \eqno{(2.6)}$$
in which $U^2\equiv\sum_{i=1}^d U_i^2$.  This implies following two results :
$$\sum_{i=1}^d  {\partial^2h^i_{~i}\over \partial U^i\partial U_i} = \sum_{i=1}^d {\partial^2\over \partial U^i\partial U_i}\left({B-C\over U^2} U_i^2\right) + \sum_{i=1}^d {\partial^2 C \over \partial U^i\partial U_i}\hspace{4cm}$$
$$ =  \sum_{i=1}^d \left[{\partial^2 \left({B-C\over U^2}\right)\over \partial U^i\partial U_i} U_i^2 + 4 {\partial \left({B-C\over U^2}\right)\over \partial U^i} U_i \right] + 2d{B-C\over U^2}  + {\vec\nabla}_U^2  C. \eqno{(2.7)}$$
$$\sum_{i\ne j}^d {\partial^2h^i_{~j}\over \partial U^i\partial U_j} = \sum_{i\ne j}^d {\partial^2\over \partial U^i\partial U_j}\left({B-C\over U^2} U_i U_j\right) \hspace{6cm}$$
$$ =  \sum_{i\ne j}^d \left[{\partial^2 \left({B-C\over U^2}\right)\over \partial U^i\partial U_j} U_i U_j \right]+ 2(d-1) \sum_{i}^d {\partial \left({B-C\over U^2}\right)\over \partial U^i} U_i + d(d-1) {B-C\over U^2}.\eqno{(2.8)}$$
Now, using the property 
$$\sum_{i}^d {\partial f\over \partial U^i} U_i = \vec U_i \cdot \vec \nabla = U{\partial f\over \partial U},\eqno{(2.9)}$$
if f=f(U), then (2.7) and (2.8) implies following simple result
$${\partial^2h^M_{~N}\over \partial x^M\partial x_N} = U^2{\partial^2 \left({B-C\over U^2}\right)\over \partial U^2}+ 2(d+1) U {\partial \left({B-C\over U^2}\right)\over \partial U} + d(d+1) {B-C\over U^2}+ {\vec\nabla}_U^2  C, \eqno{(2.10)}$$
which is a part of third term in (2.4).

$\bullet$ {\bf Step 3 :} From (2.6) we see that 
$$ h = (B-C) + d~C + \cdot\cdot\cdot.  \eqno{(2.11)}$$
Thus Eq.(2.4) tell us that $B$ and $C$  will contribute following quantity to  $ \Theta_{00}$ 
$${\partial^2h\over \partial x^Q\partial x_Q} = {\vec\nabla}^2(B-C) +d~ {\vec\nabla}^2 C\cdot\cdot\cdot ={\vec\nabla}_U^2(B-C)+({\vec\nabla}')_U^2(B-C)+ d~ {\vec\nabla}_U^2 C+d~({\vec\nabla}')_U^2 C+\cdot\cdot\cdot,  \eqno{(2.12)}$$
in which ${\vec\nabla}_U^2$  is the Laplacian on the coordinate $U_i$ while $({\vec\nabla}')_U^2$ is that on the coordinates except $U_i$.

$\bullet$ {\bf Step 4 :}  Using the formula 
$${\vec\nabla}_U^2 f(U) = {1\over U^{d-1}}\partial_U\left(U^{d-1}\partial_Uf(U)\right)=\partial_U^2 f(U) +{d-1\over U}\partial_U f(U), \eqno{(2.13)}$$
we can substitute (2.10) and (2.12) into (2.4) to find that $B$ and $C$ will contribute following into  $\kappa^2 \Theta_{00}$
$$ \kappa^2 \Theta_{00}= {1\over2}{d-1\over U^{d-1}} \partial_U\left(U^{d-1}C\right) + {1\over2} ({\vec\nabla}')_U^2 C+{1\over2}{d-1\over U^{d-1}} \partial_U\left(U^{d-2}(B-C)\right) + {1\over2} ({\vec\nabla}')_U^2(B-C)+\cdot\cdot\cdot.\eqno{(2.14)}$$
That coming from $E$ and $F$ has a similar formula after replacing with $U\rightarrow r$ and $d\rightarrow D$.

$\bullet$ {\bf Step 5 :} A simple observation form (2.4) could see that $G$  will contribute following into  $\kappa^2 \Theta_{00}$
$$ \kappa^2 \Theta_{00}=  {1\over2} d_x~({\vec\nabla}')_x^2 G+\cdot\cdot\cdot.\eqno{(2.15)}$$

Finally, using  (2.14) and (2.15) we can find the complete value of $\kappa^2 \Theta_{00}$.  After substituting it into (2.2) we then obtain the ADM mass.

\section{ADM Mass of Magnetic  Black D-brane}
The non-extremal black D-brane we considered is described by the geometry (1.3).  We express the corresponding metric in the Einstein frame as following
$$ds_{10}^2 =(1+ B^2 r^2)^{1/8} \left[H^{-3\over8}\left(-f(U)~dt^2+dz^2+dw^2+dr^2+ {r^2d\phi^2\over 1+B^2r^2}\right)+H^{5\over8} \left(f(U)~dU^2+U^2d\Omega_4^2\right) \right],\eqno{(3.1)}$$ 
$$f(U) = 1-{U_0^3\over U^3},~~~~~~~H = 1+ {U_0^3 \sinh^2\gamma \over U^3},\eqno{(3.2)}$$ 
After the calculation the ADM mass is
$$M= {1\over 2\kappa^2}\Omega_4 L_zL_w \pi R^2 U_0^3 \left[(3\sinh^2\gamma+4)\left({8\over 9}{(1+B^2R^2)^{9/8}-1\over B^2R^2}\right)\right].\eqno{(3.3)}$$ 
Here we assume that the coordinate $z$ and $w$  is compactified on the circles of circumference $L_z$ and $L_w$ respectively.  Brane also is wrapped on the radius with $0\le r \le R$.  Notice that terms which do not depend on the $U_0$ are infinite and have been dropped out from $M$, as they are that of the background and shall not be regarded as parts of the black D-brane mass.   

The temperature and entropy could be easily calculated and results are
$$ T = {3\over 4\pi U_0\cosh\gamma },\hspace{2cm}\eqno{(3.4)}$$ 
$$ S ={4\pi\over 2\kappa^2}\Omega_4 L_zL_w \pi R^2 U_0^4 \cosh\gamma,\eqno{(3.5)}$$ 
The energy denotes that above extremality is 
$$E ={1\over 2\kappa^2}\Omega_4 L_z L_w \pi R^2 U_0^3 \left[(3\sinh^2\gamma+4)\left({8\over 9}{(1+B^2R^2)^{9/8}-1\over B^2R^2}\right)-3\sinh\gamma\cosh\gamma\right],\eqno{(3.6)}$$ 
\section{Hawking-Page Phase Transition in Magnetic  Black D-brane}
To describe the dual gauge theory form above black D-brane property we have to consider the near-extremal configuration of the black D-brane.  This could be found by the following limits [7].

  First, we define $ h^3 \equiv U_0^3 \cosh\gamma\sinh\gamma$. Next, we consider the following rescaling
$$U\rightarrow{U_{old}\over \ell^2},~~~~U_0\rightarrow{(U_0)_{old}\over \ell^2},~~~~h^3\rightarrow{h^3_{old}\over \ell^2},\eqno{(4.1)}$$ 
and taking $\ell \rightarrow 0$ while keeping the old quantities fixed [7].

  In this limit we find that
$$ T = {3\over 4\pi U_0}\left({U_0^3\over h}\right)^{3/2},\hspace{7cm}\eqno{(4.2)}$$ 
$$ S ={4\pi\over 2\kappa^2}\Omega_4 L_zL_w \pi R^2 U_0^4\left({U_0^3\over h}\right)^{-3/2},\hspace{4.7cm}\eqno{(4.3)}$$ 
$$E ={1\over 2\kappa^2}\Omega_4 L_zL_w \pi R^2 U_0^3 \left[{5\over2}+{3h^3\over U_0^3}\left({8\over 9}{(1+B^2R^2)^{9/8}-1\over B^2R^2}-1\right)\right],\eqno{(4.4)}$$ 
The free energy $F=E-TS$ becomes
$$F ={1\over 2\kappa^2}\Omega_4 L_zL_w \pi R^2 U_0^3 \left[-{1\over2}+{3h^3\over U_0^3}\left({8\over 9}{(1+B^2R^2)^{9/8}-1\over B^2R^2}-1\right)\right].\eqno{(4.5)}$$ 
which becomes that in [4] when $B\rightarrow 0$ and free energy is negative.  However, for a large $B$ the free energy becomes positive. 

This means that, as noted first by Hawking and Page [10], a first order phase transition occurs at some critical temperature, above which an AdS black hole forms. On the other hand, at a lower temperature, the thermal gas in AdS dominates.   On dual gauge theory side [11],  Witten related the Hawking-Page phase 
transition of black holes in AdS space with the confinement-deconfinement phase transition of  field theory [12].  Thus we have seen that  the magnetic field could produce the Hawking-Page transition and  the corresponding dual gauge theory will show the confinement-deconfinement phase transition under large  magnetic flux.
\section{Conclusion}
In this paper we have derived a  formula which enable us to evaluated the ADM mass in more general cases.   We use  this formula to evaluate the thermodynamical quantities of the black D-branes with Melvin magnetic field, which is dual to the finite temperature gauge theory under the magnetic field. We have found the Hawking-Page transition for sufficiently large magnetic field. This means that the corresponding dual gauge theory will show the confinement-deconfinement phase transition under large magnetic flux. It is hoped that the formula derived in this paper could help us to evaluate the  ADM mass to study the thermodynamical quantities of  general black brane systems. 
\\
\\
{\bf Acknowledgments} :We are supported in part by the Taiwan National Science Council.
\newpage
\begin{center} {\bf  \Large References}\end{center}
\begin{enumerate}
\item  A. Dabholkar, G. W. Gibbons, J. A. Harvey and F. Ruiz Ruiz, Nucl. Phys. B340 (1990) 33.
\item M. J. Duff and J. X. Lu,``Black and super p-branes in diverse dimensions", Nucl. Phys. B 416, 301 (1994) [hep-th/9306052].
\item  J. X. Lu, ``ADM masses for black strings and p-branes",  Phys. Lett. B 313, 29 (1993) [hep-th/9304159].
\item  S. S. Gubser, I. R. Klebanov, A. W. Peet, ``Entropy and Temperature of Black 3-Branes", Phys.Rev.D54 (1996) 3915 [hep-th/9602135]; I. R. Klebanov and A. A. Tseytlin, ``Entropy of near extremal black p-branes", Nucl.Phys. B475 (1996) 164 [hep-th/9604089];  M.J. Duff, H. Lu, C.N. Pope, ``The Black Branes of M-theory," Phys.Lett. B382 (1996) 73 [hep-th/9604052].  
\item   Rong-Gen Cai. and Nobuyoshi Ohta, ``On the Thermodynamics of Large N Noncommutative Super Yang-Mills Theory", Phys.Rev. D61 (2000) 124012  [hep-th/9910092]; T. Harmark, N. A. Obers, ``Phase Structure of Non-Commutative Field Theories and Spinning Brane Bound States," JHEP 0003 (2000) 024  [hep-th/9911169 ].
\item  Steven S. Gubser,``Thermodynamics of spinning D3-branes," Nucl.Phys. B551 (1999) 667 [hep-th/9810225] ;  Rong-Gen Cai, Kwang-Sup Soh, ``Critical Behavior in the Rotating D-branes", Mod.Phys.Lett. A14 (1999) 1895 [hep-th/9812121].
\item T. Harmark and N.A. Obers, ``Thermodynamics of Spinning Branes and their Dual Field Theories", JHEP 0001 (2000)  008 [hep-th/9910036];  T. Harmark, V. Niarchos, N.A. Obers, ``Instabilities of Near-Extremal Smeared Branes and the Correlated Stability Conjecture," JHEP 0510 (2005) 045 [hep-th/0509011]; ``Instabilities of Black Strings and Branes",   Class. Quant. Grav. 24 (2007) R1-R90 [hep-th/0701022].
\item M.A. Melvin, ``Pure magnetic and electric geons,'' Phys. Lett. 8 (1964) 65; 
F.~Dowker, J.~P.~Gauntlett, D.~A.~Kastor and J.~Traschen, ``The decay of magnetic fields in Kaluza-Klein theory,'' Phys.\ Rev.\ D52 (1995) 6929 [hep-th/9507143]; M.~S.~Costa and M.~Gutperle, ``The Kaluza-Klein Melvin solution in M-theory,'' JHEP 0103 (2001) 027 [hep-th/0012072].
\item Wung-Hong Huang, ``Holographic Gauge Theory with Maxwell Magnetic Field," [arXiv:0904.2328v5 [hep-th]];  `` Semiclassical Strings in Electric and Magnetic Fields Deformed $AdS_5 \times S^5$ Spacetimes," Phys.Rev.D73 (2006) 026007 [hep-th/0512117 ]; ``Spin Chain with Magnetic Field and Spinning String in Magnetic Field Background," Phys.Rev. D74 (2006) 027901 hep-th/0605242; ``Giant Magnons under NS-NS and Melvin Fields," JHEP0612 (2006) 040 [hep-th/0607161].  
\item S. W. Hawking and D. N. Page,``Thermodynamics Of Black Holes In Anti-De Sitter Space," Commun. Math. Phys. 87, 577 (1983).
\item J.~M. Maldacena, ``The large N limit of superconformal field theories  and supergravity,''  Adv. Theor. Math. Phys.  2  (1998) 231-252  [hep-th/9711200]; E.~Witten, ``Anti-de Sitter space and holography,'' Adv.\ Theor.\ Math.\ Phys.\   2 (1998) 253 [hep-th/9802150].
\item E.~Witten, ``Anti-de Sitter space, thermal phase transition, and confinement in  gauge theories,'' Adv.\ Theor.\ Math.\ Phys.\   2 (1998) 505 [hep-th/9803131]. 
[hep-th/9802109].
\end{enumerate}
\end{document}